\documentclass[manuscript]{acmart}

\usepackage{float}
\usepackage{subfigure}
\AtBeginDocument{%
  \providecommand\BibTeX{{%
    \normalfont B\kern-0.5em{\scshape i\kern-0.25em b}\kern-0.8em\TeX}}}
\usepackage{tabularx}
\usepackage{bbding}
\usepackage{graphicx}
\usepackage{geometry}
\usepackage{subfigure}
\usepackage{amsmath}
\usepackage{makecell}
\usepackage{float}
\usepackage{color}
\usepackage{booktabs}
\usepackage{caption}
\usepackage{tabu}
\usepackage{hyperref}

\copyrightyear{2025}
\acmYear{2025}
\setcopyright{acmlicensed}\acmConference[CHI '25]{Proceedings of the 2025 CHI Conference on Human Factors in Computing Systems}{April 26-- May 1, 2025}{Yokohama, Japan}
\acmBooktitle{Proceedings of the 2025 CHI Conference on Human Factors in Computing Systems (CHI '25), April 26-- May 1, 2025, Yokohama, Japan}
\acmDOI{XXXXXXX.XXXXXXX}

\author{Hongxi Pu$^{*}$}
\email{hongxi@umich.edu}

\affiliation{
  \institution{University of Michigan}
  \city{Ann Arbor}
  \country{United States}
}

\author{Futian Jiang}

\affiliation{
  \institution{University of Hong Kong}
  \city{Hong Kong}
  \country{Hong Kong}
}

\author{Zihao Chen}

\affiliation{
  \institution{Brooklyn College}
  \city{NYC}
  \country{United States}
}

\author{Xingyue Song}

\affiliation{
  \institution{San Francisco Conservatory of Music}
  \city{San Francisco}
  \country{United States}
}

\begin{document}
\title[ComposeOn Academy]{ComposeOn Academy: Transforming Melodic Ideas into Complete Compositions Integrating Music Learning}

\begin{abstract}
Music composition has long been recognized as a significant art form. However, existing digital audio workstations and music production software often present high entry barriers for users lacking formal musical training. To address this, we introduce ComposeOn, a music theory-based tool designed for users with limited musical knowledge. ComposeOn enables users to easily extend their melodic ideas into complete compositions and offers simple editing features. By integrating music theory, it explains music creation at beginner, intermediate, and advanced levels. Our user study (N=10) compared ComposeOn with the baseline method, Suno AI, demonstrating that ComposeOn provides a more accessible and enjoyable composing and learning experience for individuals with limited musical skills. ComposeOn bridges the gap between theory and practice, offering an innovative solution as both a composition aid and music education platform. The study also explores the differences between theory-based music creation and generative music, highlighting the former's advantages in personal expression and learning.
\end{abstract}



\begin{teaserfigure}
\centering
\includegraphics[width=0.5\linewidth]{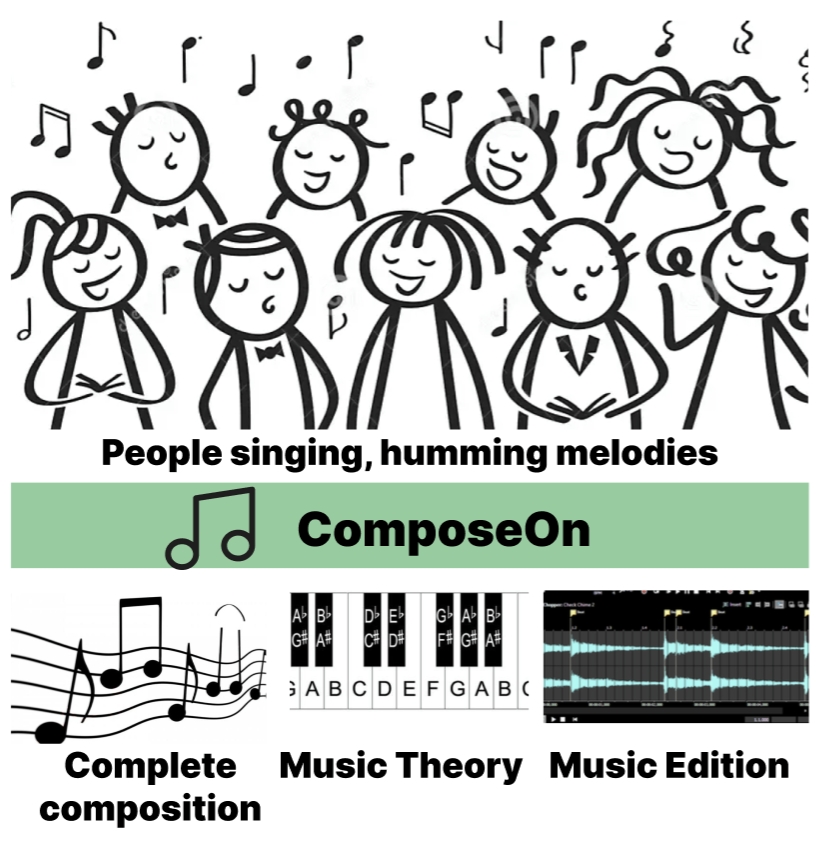}
\caption{ComposeOn takes the melody input to compose complete music, teach music theory and allow music edition.}
\Description{ComposeOn System Design Diagram with three modules}
\label{fig:system_design_diagram}
\end{teaserfigure}

\maketitle

\section{Introduction}\label{sec:Introduction}
Music composition has long been recognized as a significant and widely appreciated art form. With the advent of the digital age, there has been a growing desire among individuals lacking formal musical training to express themselves through music creation. This demand has driven the development of various digital audio workstations (DAWs) and music production software. However, despite technological advancements, many existing tools still present substantial barriers to entry and fail to adequately address the needs of novice users.

Current systems and software often fall short of providing sufficient support for novice users in music composition. Traditional DAWs, while powerful, can be overwhelmingly complex for learners with limited musical skills ~\cite{r16}. These tools typically require users to possess a solid foundation in music theory, familiarity with musical notation, and proficiency in navigating intricate software interfaces. This complexity gives rise to three primary challenges. Firstly, the high technical threshold makes it difficult for beginners to master these tools ~\cite{r14}. Secondly, existing tools often struggle to effectively translate novice users' creative ideas into musical compositions  ~\cite{r15}. Lastly, these software applications frequently fail to effectively impart additional music theory knowledge to users ~\cite{r17}. Collectively, these factors contribute to a significant gap between individuals' desire to create music and their actual ability to do so, particularly for those without formal musical training ~\cite{r13}. This disparity not only impedes personal creative expression but may also potentially diminish the diversity and innovation in music creation  ~\cite{r21}. Consequently, the development of more intuitive, user-friendly, and learning-supportive music creation tools has emerged as a crucial research direction  ~\cite{r16}.

To address this challenge, we present ComposeOn, a music theory-based tool designed specifically for users with little to no music knowledge. ComposeOn empowers users to easily extend and develop their melodic ideas into complete compositions. Additionally, the tool offers easy editing, which allows users to effortlessly edit and control their music creations. By integrating music theory, ComposeOn explains the music creation from the music theory perspectives in three different levels: beginner, intermediate, and advanced.

Our user study (N=10) compared participants composing music using ComposeOn and the baseline method, Suno AI, to explore the effectiveness of ComposeOn in helping novice users create music. The results show that ComposeOn provides a more accessible and enjoyable composing and learning experience for individuals with little musical skill. 

In our discussion, we explored how ComposeOn leverages music theory to enhance music creation for beginners. Unlike generative AI tools, ComposeOn provides a structured, educational approach, encouraging users to actively engage with musical concepts. The tool's melody input feature allows users to develop their ideas into complete compositions, fostering creativity and personal expression. Additionally, ComposeOn's multi-tiered explainability enhances learning by offering tailored insights into music theory, promoting deeper understanding and critical thinking. This innovative approach positions ComposeOn as both a composition aid and a powerful music education platform, bridging the gap between theory and practice.

This study highlights the contributions of ComposeOn in addressing user dissatisfaction with generative music tools. Our user study findings indicate that many users find generative music lacking in coherence and personal expression. ComposeOn bridges this gap by integrating music theory with practical composition, allowing users to engage deeply with the creative process. The tool's focus on melody input and explainability supports users in developing their musical ideas while enhancing their understanding of music theory. This approach not only improves user satisfaction but also fosters a more meaningful music creation experience.

\section{Related Work}\label{sec:Related Work}
\subsection{Compose Theory}
ComposeOn promotes people without a music background to compose their own music. We design the system to detect melody users put in and give suggestions for continuations based on basic music and harmony progression theory. "Western music written during Baroque, Classical, and Romantic periods (ca.1650-ca.1900) is called tonal music, which has a point of gravitation called tonic." \citep{laitz2008complete} The keys and scales gradually formed based on that and thus formed tonal hierarchy and harmony function theory.

"The music of the tonal era is almost exclusively tertian, which means being constructed of stacked 3rd." \citep{kostka2006materials} Chords are marked with their root notes (where the stack begins) using roman numerals. Functionally, they are basically divided into Tonic chords: I, Dominant chord: V, and predominant/subdominant chords: ii, iv, vi. The harmony often progresses as Tonic–Subdominant–Dominant–Tonic. Sometimes may use iii or Vii, and different 7th chords, but their function is various depending on the texture \citep{aldwell2010harmony}.

There are also uses out of this basic progression, such as sequence, modulation or transportation. We also considered those situations with limited possibilities within our database.

Based on the above theory, we suggest notes within the key and the possible harmony progressions. As to music phrases, we followed basic 2+2/4+4 (refer to measure numbers) to form the music phrase and thus sentence and sections. \citep{schoenberg1967fundamentals} We mainly focus on songwriting, so the intro, verse — chorus — verse — chorus —bridge — chorus — outro structure of song is also being considered. \citep{masterclass2021songwriting}

\subsection{Voice to MIDI Technology and Its Applications}

MIDI (Musical Instrument Digital Interface) is a technical standard that describes a protocol, digital interface, and connectors, allowing various electronic musical instruments, computers, and other related devices to connect and communicate with each other \citep{midi1996complete}. Voice to MIDI technology is the process of converting vocal or other audio signals into this MIDI data format, playing a crucial role in music production and analysis. This technology involves multiple steps, including pitch detection, note segmentation, quantization, and MIDI conversion. Pitch detection typically employs algorithms such as YIN \citep{decheveigne2002yin} or pYIN \citep{mauch2014pyin} to estimate the fundamental frequency of audio. Subsequently, the continuous pitch sequence is segmented into discrete notes, which are then time-aligned to a musical grid. Finally, the detected note information is converted into MIDI events. Voice-to-MIDI technology has found applications in various fields, such as quickly converting hummed melodies into MIDI for song composition, providing instant feedback to students in music education, and generating real-time music based on user voice input in interactive music systems. This technology not only simplifies the music creation process but also provides powerful tools for music analysis and education.

\subsection{Automated Melody Analysis}

Automated melody analysis is a significant branch of music information retrieval, aimed at extracting and analyzing melodic features from musical data. This process typically includes the analysis of notes, chords, and chord progressions. Note analysis involves extracting attributes such as pitch, duration, and velocity. Chord analysis focuses on identifying combinations of simultaneously sounding notes, often using algorithms like Chordino \citep{mauch2010simultaneous}. Chord progression analysis examines patterns in the series of chords, utilizing methods such as hidden Markov models \citep{rohrmeier2012comparing}. In this field, Musicpy, a powerful Python music programming library, provides extensive functional support. It not only has concise syntax for representing various musical elements but also incorporates a complete music theory system supporting advanced musical operations. Musicpy's core data structures include notes, keys, chords, scales, etc., offering various practical functions including chord identification, melody analysis, and automatic composition. Through Musicpy, researchers and music creators can conveniently achieve automated melody analysis, explore musical structures and characteristics. Notably, the ComposeOn project extensively utilizes Musicpy's powerful capabilities, particularly in chord extraction and chord progression analysis. ComposeOn employs Musicpy's algorithms to identify and analyze the progression patterns of these chords, thereby providing an important foundation for music analysis and creation. This application demonstrates the practicality and effectiveness of Musicpy in real-world music analysis projects.

\subsection{Automatic Accompaniment Generation}
Accompaniment generation is referred to as "the audio realization of a chord sequence"
by systems like MySong \cite{simon2008mysong}, which represents a significant advancement in the field of automatic accompaniment generation for vocal melodies. MySong allows users to input vocal melodies, which the system then inputs to a hidden Markov model to recommend chords. However, MySong's capabilities are limited to generating accompaniments, whereas our ComposeOn system empowers users to easily extend and develop their melodic ideas into complete compositions, providing a more comprehensive music creation and learning experience.

\section{Formative Study}\label{sec:Formative Study}
To understand the needs and challenges faced by individuals with little to no music theory knowledge in composing music, we conducted a formative study with six participants (FP1-FP6), aiming at exploring their willingness and confidence in music composition, as well as their experiences with existing music composition tools, the demographics and basic music composition information are shown in \ref{tab:music_composition}.
\begin{table}[h]
\centering
\resizebox{\textwidth}{!}{%
\begin{tabular}{|c|c|c|c|p{5cm}|}
\hline
\textbf{Labels} & \textbf{Music Theory Level} & \textbf{Compose Willingness} & \textbf{Compose Confidence} & \textbf{Compose Tools} \\
\hline
FP1 & Intermediate & Moderate & Moderate & Suno AI \\

FP2 & Beginner & High & Low & GarageBand, Suno AI \\

FP3 & Beginner & High & Low & Logic Pro, Fruity Loops \\

FP4 & Intermediate & High & Moderate & Suno AI \\

FP5 & Beginner & Moderate & Low & Never \\

FP6 & Intermediate & High & Low & Udio \\
\hline
\end{tabular}%
}
\caption{Music Composition Profile of Participants}
\label{tab:music_composition}
\end{table}

Most participants expressed a strong desire to compose music, with many having attempted to use various music composition tools in the past. However, they encountered significant barriers due to the tools' requirements for \textbf{basic music theory knowledge}. For instance, FP2 mentioned, \textit{"I tried GarageBand~\cite{r1}, but I struggled with identifying piano keys and understanding orchestration techniques".} Other composition tools, such as Logic Pro~\cite{r2} ~\cite{r4}and Fruity Loops Studio~~\cite{r3}~\cite{r5}, were also cited as challenging for beginners due to their requirements for music theory knowledge.

Recent developments in AI-powered music generation, such as Udio~\cite{r8} and Suno AI~\cite{r7}~\cite{r6}, were acknowledged by participants as interesting music composition tools. However, users expressed concerns about the \textbf{low interpretability} and \textbf{limited control} in AI music generation, leading to a sense of disconnection from the creative process. FP2 remarked, \textit{"It's fascinating to see what the AI can produce, but I often feel like I'm just pressing buttons rather than truly composing."} Rather than feeling like they were composing music themselves, participants viewed these tools more as a form of entertainment or game. FP1 elaborated on this sentiment, stating, \textit{"It's fun to play around with, but I don't feel like I'm learning or improving my musical skills."} These characteristics also \textbf{hindered users' from learning music theory} from AI-generated compositions and did not increase their their own music composition confidence. FP4 noted, \textit{"Suno can generate and extend music, but I feel limited in my ability to adjust the output or incorporate my own creative ideas".} This sentiment highlights a gap between the capabilities of AI-generated music and the desire for personal creative input. FP6 further emphasized this point, saying, \textit{"I want to understand why certain musical choices are made, but the AI doesn't provide that insight. It's like being given a finished painting without learning how to mix colors or use brushstrokes."} Several other participants echoed similar concerns, emphasizing the importance of maintaining creative agency and the ability to learn and grow as musicians through the composition process.

Despite their lack of formal music theory knowledge, all participants reported frequently experiencing melodic inspirations. FP5 expressed, \textit{"I often have tunes in my head that I'd love to develop into full songs, but I don't know where to start".} This desire to \textbf{expand on their melodic ideas and potentially create complete songs} was a common theme among participants. However, the participants generally lacked the motivation to undertake extensive music theory study, from understanding notes to learning chord progressions, as a prerequisite to composition. FP3 stated, \textit{"The idea of studying music theory from scratch before I can start composing is daunting and discouraging".} Interestingly, when presented with the concept of learning music theory gradually through the composition process, participants showed increased enthusiasm. FP5 commented, \textit{"If I could learn about music while actually creating something, I'd be much more motivated to stick with it".}

This formative study revealed a clear need for a music composition tool that caters to beginners or intermediate music theory learners, allowing them to \textbf{translate their melodic ideas into compositions} without extensive prior knowledge. Additionally, the generated music should be \textbf{easy for users to edit}, giving them control over their creations. Furthermore, it highlighted the potential value of integrating \textbf{music theory learning} into the composition process itself, potentially increasing user engagement and long-term commitment to music creation.

\section{System Design}\label{sec:System Design}
Our system, ComposeOn, is designed to facilitate music extension and learning for users with little or no musical theory background. By analyzing the results of the formative study mentioned in Section \ref{sec:Formative Study}, we identified three design goals for ComposeOn:

\begin{itemize}
\item Melody Expansion: Enable users to easily extend and develop their melodic ideas into complete compositions.
\item Easy Editing: Allow users to effortlessly edit and control their music creations.
\item Music Theory Integration: Incorporate music theory learning to boost engagement and interest.
\end{itemize}

To generate the extended part of music based on the input melody, our system consists of three main components: the input \& analysis module, the generation module, and the output \& explanation module. Figure \ref{fig:system_design_diagram} shows a high-level diagram of the system design.
\begin{figure}[h]
\centering
\includegraphics[width=0.8\linewidth]{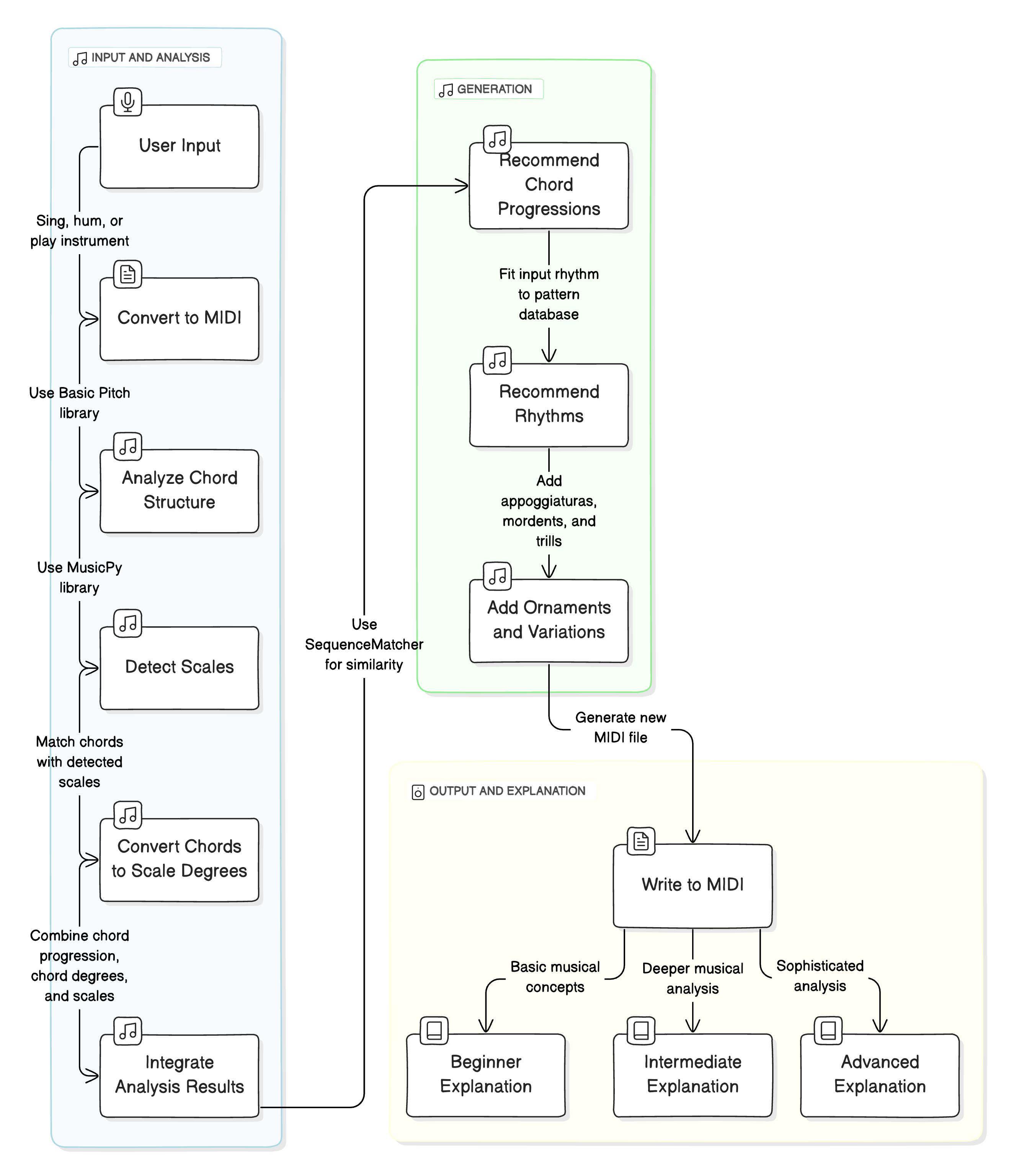}
\caption{ComposeOn System Design Diagram: the input and analysis module, represented by light-blue color; the generation module, represented by light-green color; and the output and explanation module, represented by light-yellow color.}
\Description{ComposeOn System Design Diagram with three modules}
\label{fig:system_design_diagram}
\end{figure}

\subsection{ComposeOn Database}

The ComposeOn database consists of two main parts: common chord progressions and common rhythm patterns. These data provide ComposeOn with fundamental materials and references.

The \textbf{chord progression section} contains 39 different chord sequences across seven categories, covering various music styles from basic to advanced. These chord progressions are selected based on the principles of Functional Harmony and Tonal Harmony in music theory\cite{kostka2008harmonia,clendinning2016musician}. Specifically, it includes (1) 9 classic chord progressions, such as \texttt{I–IV–V–I} and \texttt{vi–IV–V–I}, common in pop and rock music; (2) 9 extended chord progressions, like \texttt{Imaj7–ii7–V7–Imaj7}, suitable for jazz and blues; (3) 4 diminished triad progressions, such as \texttt{i–iidim–V7–i}, used to create tense or dissonant effects; (4) 4 augmented fourth chord progressions, like \texttt{I–IV–aug4–I}, used to add harmonic color; (5) 5 mixed chord progressions, such as \texttt{Imaj7–ii7–V7–IVmaj7}, blending different types of chords; (6) 4 substitute chord progressions, like \texttt{Imaj7–bIImaj7–V7–Imaj7}, common in modern jazz; and (7) 4 cycle chord progressions, such as \texttt{Imaj7–ii7–V7–iii7}, used to create repeating sections or build-ups. The selection of these chord progressions also references popular music composition practices and jazz theory\cite{levine2011jazz,mulholland2013berklee}.

The \textbf{rhythm pattern section} contains 16 different rhythm types, each composed of specific combinations of notes and rests~\cite{r9}. These patterns cover various popular music styles, including pop, rock, reggae, ska, jazz, funk, blues, and country music. For example, the first pattern \texttt{[(1, 'rest'), (1, 'note'), (1, 'rest'), (1, 'note')]} represents a simple, regular rhythm, while the seventh pattern \texttt{[(1/3, 'note'), (1/3, 'note'), (1/3, 'note')] * 4} represents a more complex, triplet-based rhythm. These patterns provide ComposeOn with a rich selection of rhythms, enabling it to generate melodies that conform to specific musical styles.

\subsection{Input and Analysis Module}

To allow users to easily express their musical ideas, we have designed a flexible input module. Users can input their melodies by singing, humming, or playing an instrument. To ensure an accurate capture of users' musical creativity, we employ the advanced Basic Pitch library~\cite{r10}. This neural network-based pitch detection model can accurately convert users' audio input into standard MIDI file format, handling even complex polyphonic melodies with precision.

Once the user's input is converted into a MIDI file, the analysis module begins its work, delving deep into the musical characteristics of the melody. We use the powerful MusicPy~\cite{r11} library to process this MIDI file, extracting rich musical information. First, ComposeOn identifies the implied \textbf{chord progressions} in the melody, for example, the chords present in the user's input, recognizing them as D and G.. This provides insight into the harmonic foundation of the melody. Next, ComposeOn uses the detect\_scale function to determine the \textbf{scales} most likely used in the melody, for example, the system determines that the scale being used is D major.. To standardize the melody's harmonic structure, ComposeOn converts the identified chords into \textbf{scale degrees}, within the context of D major: D is identified as the I (tonic) chord, and G is identified as the IV (subdominant) chord. This process entails matching chords with detected scales, determining their position within the scale, and addressing special cases for minor scales. This conversion provides a more abstract, theoretical representation of the harmonic structure, allowing for a unified generation logic to be applied regardless of the original scale or key.

Finally, ComposeOn integrates all these analysis results, creating a detailed profile of the input melody's characteristics, and serving as a basis of the explanations. The standardized scale degree representation enables ComposeOn to apply consistent generation algorithms across various musical contexts.

\subsection{Generation Module}

The Generation Module is responsible for creating new musical content based on the analysis of the user's input. This process is divided into three main steps: recommending chord progressions, recommending rhythms for right-hand rhythm, and adding ornaments and variations. The system consults its chord progression database and identifies a common progression pattern that incorporates the input chords: [I - IV - V - I];[I - IV - ii - V - I];  [I - IV - vii dim - I];  [I - IV - V7 - I] etc.

\textbf{Step 1: Recommending Chord Progressions}

In the first step, we begin with chord degrees from the previous step of the Analysis Module and find the recommended chord degrees. When the user clicks the "Continue" button in the UI, we sequentially identify the most similar progressions, utilizing the SequenceMatcher class from the difflib library ~\cite{pythondifflib} to calculate the sequence similarity between the input chord progression and our predefined progressions through their chord degrees.  Each click recommends a complete progression. This recommended progression is then combined with the original input progression to form a new progression base. On subsequent "continue" clicks, this combined progression serves as the new input. 

Then, we convert these degrees into absolute chords based on the scales retained from earlier steps of the Analysis Module. For instance, if our recommended chord degree sequence is [I, IV, V, I], and the retained scale is D major, the resulting chord progression would be [D, G, A, D]. This determined chord progression forms the basis for both the left-hand and right-hand melodies.

For the left-hand part, we utilize the triads of these chords, playing them as whole notes for each measure. In the case of [D, G, A, D], the left hand would play D-F\#-A (D major triad) for the first measure, G-B-D (G major triad) for the second measure, A-C\#-E (A major triad) for the third measure, and D-F\#-A (D major triad) for the fourth measure. This approach provides a solid harmonic foundation using simple, sustained chords in the left hand.

Meanwhile, the same chord progression serves as the foundation for developing the right-hand melody, allowing for more intricate melodic and rhythmic patterns that complement the underlying harmonic structure. For example, the right-hand melody might incorporate arpeggios or scalar passages derived from the D major scale, with emphasis on the chord tones of each underlying harmony.

This method of chord realization and melody generation demonstrates how the system can translate abstract music theory concepts into concrete musical elements. By providing a clear harmonic foundation in the left hand and a related but more elaborate melody in the right hand, the system creates a balanced and musically coherent output that is accessible to novice users while still adhering to established musical principles.

Furthermore, this approach can be extended to the other progression options identified earlier:

\begin{enumerate}
    \item For [I, IV, ii, V, I] in D major: [D, G, Em, A, D]
    \item For [I, IV, vii°, I] in D major: [D, G, C\# dim, D]
    \item For [I, IV, V7, I] in D major: [D, G, A7, D]
\end{enumerate}

To introduce uniqueness, we then apply variations to these selected progressions. This variation technique is based on the concept of chord substitution in music theory. According to Levine (2011), chords can often be substituted with chords that share common tones or have a similar function within the key\cite{levine2011jazz}. Our implementation focuses on diatonic substitutions, where chords are replaced by others from the same key, maintaining harmonic coherence while introducing variety. It's worth noting that we treat each complete chord progression as a musical phrase, providing a structural basis for subsequent melody generation.

\textbf{Step 2: Recommending Rhythms for the right-hand melody}

In the second step, we employ the rhythm pattern pool introduced in the Section of the ComposeOn database. We begin by fitting the input rhythm to this rhythm pattern database to find the closest match and randomly chose two more patterns in our rhythm pattern pool. For each complete chord progression, we then apply the following strategy for rhythm patterns: The first measure of each phrase always uses the rhythm pattern fitted to the input. This ensures that the generated music retains the rhythmic characteristics of the original input, maintaining musical coherence. For subsequent measures within the phrase, we select the randomly chosen two rhythm patterns. This approach both maintains a connection to the original input and introduces new variations, enhancing the richness and diversity of the music.

\textbf{Step 3: Adding Ornaments and Variations}

In this final step, we enrich the melody by adding musical ornaments and variations. We focus on adding ornaments to only the right-hand melody, randomly selecting 5\% of the notes for embellishment. The ornaments are chosen to be as close as possible to the chord tones of the current triad. This approach includes ornaments such as appoggiaturas (grace notes creating brief dissonance before resolving to the main note), mordents (rapid alternations between the main note and an adjacent note), and trills (quick alternations between two adjacent notes)~\cite{adler1989study}. To implement this, we first determine the total number of notes in the right-hand melody and calculate 5\% of this total to decide how many notes will receive ornaments. We then randomly select these positions within the melody. For each chosen note, we identify the nearest chord tone based on the current harmony and select an appropriate ornament type. When adding the ornaments, we ensure they complement the harmonic structure and maintain the overall flow of the melody. This method effectively increases the expressiveness and complexity of the melody while preserving its essential character and harmonic integrity. Care is taken to use ornaments judiciously, avoiding overuse that might disrupt the melody's fluency, and to ensure their application aligns with the specific musical style and period being emulated.

By combining these three steps - chord progression generation with variations, flexible rhythm pattern application, and ornament addition - our Generation Module creates musically rich and varied content based on the user's input. This multi-faceted approach ensures that the generated music is harmonically sound, rhythmically interesting with a balance of familiarity and novelty, and melodically expressive. This provides users with inspiring and unique musical ideas that both respect the original input and introduce new musical elements.
\subsection{Output and Explanation Module}

The output module writes the generated melody to a new MIDI file, which can be played back to the user or further edited and refined. The explanation component of this module provides insights into the composition at three levels of complexity: Beginner, Intermediate, and Advanced. These explanations focus on three key aspects of the generated melody: chord progressions, rhythm patterns, and embellishments.

\textbf{Beginner Level} The explanation starts with an introduction to basic musical concepts. For chord progressions, it introduces the concept of chords as groups of notes played together, explaining the difference between major and minor chords, and showing how the melody notes relate to these underlying chords. The rhythm explanation at this level covers basic note durations such as quarter notes and eighth notes, and introduces common time signatures like 4/4 and 3/4. It demonstrates how the melody's rhythm fits into these basic patterns. Regarding embellishments, the beginner explanation introduces the concept as extra notes that decorate the main melody, providing simple examples like grace notes or trills.

\textbf{Intermediate Level} This level deepens the musical analysis. For chord progressions, it explains common sequences (e.g., I-IV-V-I) and introduces the concept of harmonic function (tonic, dominant, subdominant), discussing how the chosen progression supports the melody. The rhythm explanation at this level delves into how the input rhythm was matched to one of the 16 predefined patterns and how the other three random variations were created. It also discusses how these rhythms relate to different musical styles. The embellishment explanation introduces more complex decorative techniques like arpeggios or turns, and explains how these relate to the underlying harmony.

\textbf{Advanced Level} At this level, the explanation provides a sophisticated analysis of the composition. The chord progression explanation discusses advanced harmonic concepts such as secondary dominants or modal interchange, explains any modulations or key changes, and analyzes how the chord progression contributes to the overall structure of the piece. The rhythm explanation covers complex concepts like syncopation or polyrhythms, explaining how the rhythm interacts with the harmonic rhythm and contributes to the overall feel or genre of the piece. For embellishments, the advanced explanation discusses techniques like counterpoint or voice leading, explaining how embellishments can create tension and release in the melody and how they contribute to the overall expressiveness of the piece.

What's more, ComposeOn also incorporates a \textbf{Music Theory Mentor Chatbot}, powered by the advanced ChatGPT-4 model, to enhance the user's learning experience and provide on-demand musical expertise. Within the recommendation rationales provided by the system, specialized musical terminology is hyperlinked. When a user clicks on one of these hyperlinked terms, the query is automatically populated in the Music Theory Mentor's input field, situated in a dedicated section of the interface. This mechanism facilitates immediate access to additional information, enabling users to explore complex musical concepts without disrupting their creative flow.

\subsection{ComposeOn User Interface (UI)}
\begin{figure}[h]
\centering
\includegraphics[width=0.8\linewidth]{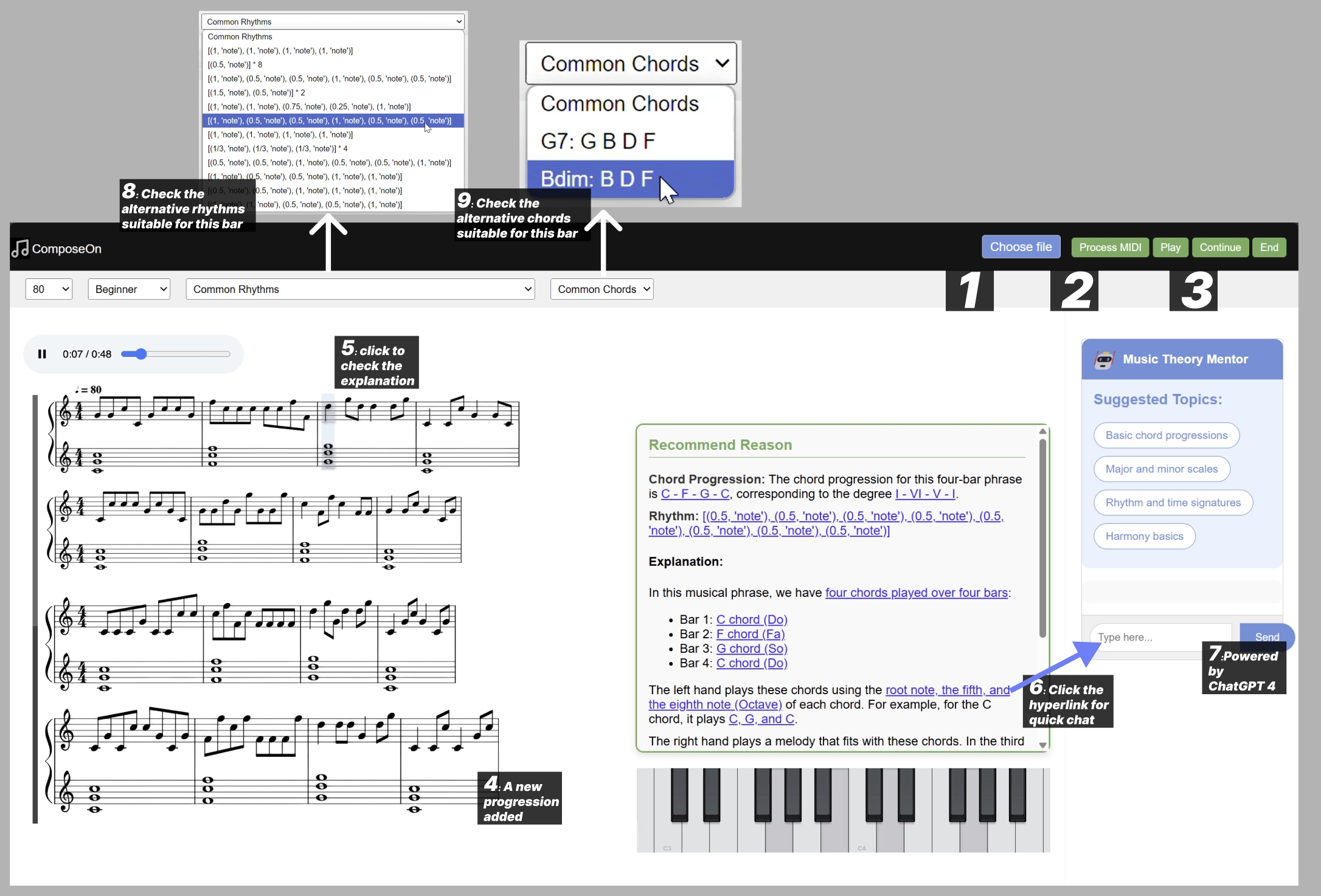}
\caption{ComposeOn UI illustration. Step1-2, choose a file, and process the file as MIDI. Step3-4, when click on the continue button, a new progression will be added. Step 5-6, click one bar or some notes to check the explanation. Step 7, click the hyperlink for quick chat with a chatbot powered by ChatGPT 4. Step8-9, check the alternative rhythms and chords suitable for the selected bar.}
\Description{ComposeOn System Design Diagram with three modules}
\label{fig:musicUI}
\end{figure}

The ComposeOn UI, as shown in Figure~\ref{fig:musicUI}  is a user-centric platform designed to facilitate seamless interaction between users and the ComposeOn composition system. This interface integrates melody continuation, editing, and explanatory functionalities, providing users with a comprehensive music creation and learning environment.

At the top of the interface, users can dynamically adjust two primary parameters: Beats Per Minute (BPM) and Explanation Level. These controls allow real-time modification of playback speed and the depth of musical analysis provided, respectively.

The composition process begins with the user uploading an audio file in either .mp3 or .wav format, initiating the Upload Module. Subsequently, activating the "Process MIDI" function triggers the Analysis Module, which visualizes the uploaded melody as a musical score on the interface. This score is interactive, allowing playback with synchronized highlighting of the current measure. Concurrently, a virtual piano display in the lower left corner visually represents the notes being played.

Upon user initiation of the "Continue" function, the Generate Module extends the composition, producing a complete chord progression. This extension is seamlessly integrated into the existing musical score visualization. When a user selects any measure of the generated continuation, the system displays explanations calibrated to the preset Explanation Level. These explanations feature hyperlinked musical terminology, enabling users to access more detailed information in the Music Theory Mentor section on the right side of the interface.

For users wishing to edit the generated music, the interface offers measure-specific editing capabilities. By selecting a measure, users can access dropdown menus for "Common Progression Degrees" and "Common Rhythms". These menus present contextually appropriate chord progression degrees and rhythmic patterns, respectively. As users modify these parameters, the system dynamically updates both the musical score and the accompanying explanations, providing immediate feedback on the musical implications of their choices.

\section{Evaluation}\label{sec:Evaluation}
A user study was conducted to evaluate the effectiveness and educational value of ComposeOn. The purpose of the study was to compare the quality and experience of two groups of people - those with no knowledge of music theory and those with some knowledge of music theory - when using ComposeOn and a benchmark method (the Suno music continuation feature) for English lyrics continuation. We paid particular attention to changes in participants' knowledge of music theory. This study aims to answer the following question:
\textbf{Q1:} Does ComposeOn \textbf{\textit{develop}} music better than Suno?
\textbf{Q2:} Does ComposeOn increase more \textbf{\textit{willingness and confidence}} of participants to develop and compose music? 

To better collect and validate the results, we had participants fill out both a pre-study and a post-study questionnaire. the pre-study questionnaire included their demographic information, a simple test of their level of knowledge of music theory, as well as their confidence and motivation about composing and learning to compose. the post-study questionnaire included their judgment of the quality of the continued music, as well as a few indicators of their judgment in SUS ~\cite{r23}. The post-study questionnaire included a quality rating of the continued music, as well as SUS indicator questions, a music theory questionnaire similar to the pre-study questionnaire, and a change in their motivation and confidence about composing.

\subsection{Participant}

\begin{table}[h]
\centering
\begin{tabular}{|c|c|c|c|}
\hline
Participant ID & Music Theory Level & Compose Freq. & Compose Willingness \\
\hline
1 & Beginner & Never & High \\
2 & Intermediate & Rarely & Medium \\
3 & Advanced & Weekly & Very High \\
4 & Beginner & Monthly & Low \\
5 & Intermediate & Daily & High \\
6 & Beginner & Yearly & Medium \\
7 & Advanced & Weekly & High \\
8 & Intermediate & Never & Low \\
9 & Beginner & Rarely & Very High \\
10 & Advanced & Daily & Medium \\
\hline
\end{tabular}
\caption{Participant Information on Music Theory and Composition}
\label{tab:musicinfo}
\end{table}

\subsection{Procedure}
\textbf{Pre-Study Questionnaire}
A pre-study questionnaire is a survey that is used prior to the start of a study or program to gather background information and initial musical knowledge about the participant. The questionnaire usually contains the following questions: \textbf{Basic information}: e.g., name, age, etc. \textbf{Relevant experience}: e.g. previous experience in music making. \textbf{Assessment of prior knowledge}: Tests the participant's current knowledge of the research topic, such as music theory. \textbf{Interest and Motivation}: To find out the participants' level of interest in the topic and their motivation to learn. \textbf{Self-efficacy}: To assess the participant's confidence in his/her ability in the domain.

\textbf{Main Study}
is to have the participants randomly pick one of the 9 melodies we prepared, and then have the participants continue the melody using the ComposeOn and baseline methods. Our baseline is the Suno 3.5 model, participants need to upload the melody file, and Suno will generate at most 4 minutes of continuation. In addition, participants can use ComposeOn to continue the melody, there is no time limit requirement, and users can check the reason for continuing the melody during the process of continuing the melody, make changes to the melody, check the knowledge of music theory through the music theory mentor, and so on.

\textbf{Post-Study Questionnaire}
The post-study questionnaire contained the same \textbf{music theory questions} as the pre-study questionnaire, which was designed to test whether the user's knowledge of music theory increased after using ComposeOn and suno for melodic continuation. In addition, there are questions about the ability of the two instruments to increase compositional \textbf{confidence and motivation}, as well as questions about the use of ComposeOn in the \textbf{SUS framework}, which are about user experience.

\section{Result}\label{sec:Result}
\begin{figure}[h]
\centering
\includegraphics[width=0.8\linewidth]{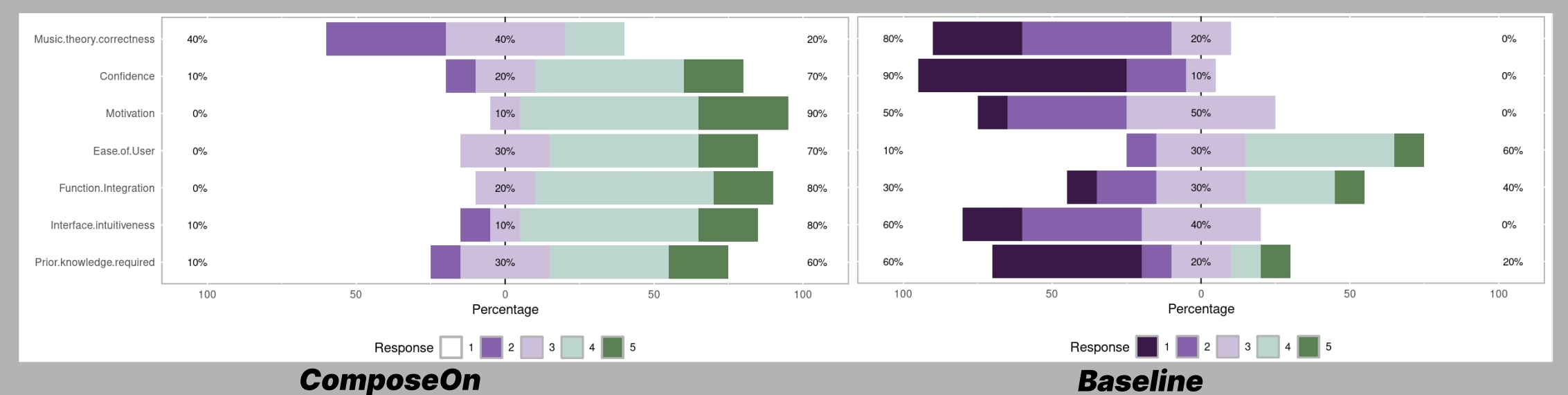}
\caption{5-point result on the music theory correctness and the subjective assessments.}
\label{fig:result}
\end{figure}

In the post-quesionnaire, we analyzed the music theory correctness and all the subjective assessments in the form of 5-point scores and free comments, Figure~\ref{fig:result} reflects the results of scores for ComposeOn and the baseline method.

\subsection{ComposeOn develops better music than Suno}
First, the music generated by Suno exhibits more richness and complexity in terms of weaving and orchestration. As noted in P2, P4, and P10, this complexity makes Suno's music potentially more appealing on first listen. One participant might describe it this way, “Suno's music feels rich on first listen, with multiple layers of sound and rich orchestration, giving a first impression of great depth. However, after multiple listens, Suno's music feels less logical. In contrast, ComposeOn's music, although given only melodies with no instrumental layers, feels comfortable after multiple listens because it is very logical.”

However, ComposeOn received high ratings for other aspects of music quality. Multiple participants (P1, P2, P9) emphasized the better structure of the music generated by ComposeOn.P1's comment was particularly specific: “The music generated by Suno lacks structure and sounds like randomly assembled pieces. It sounds like randomly assembled fragments because it has no clear development or ending. In contrast, I could hear (and see) the beginning and end of each phrase of ComposeOn's music, and each phrase ended in a similar pattern, making the music feel more holistic.”

In terms of musical coherence, several participants (P3, P6, P8) noted that ComposeOn showed better consistency in musical sequences.P3's observation was particularly insightful: “Sometimes Suno repeats chords that have been entered previously, but when you hear certain parts of the music, Suno sometimes generates random segments that have little to do with the previous text. This interspersing makes for very little musical consistency in this continuation. By contrast, the musical consistency in ComposeOn's continuation is stable; instead of outputting all the input melody at once and writing it all by itself at once, he assigns features of the input melody to different phrases, so that each phrase has parts that relate to the input melody but are not entirely consistent with it.” This approach makes the music generated by ComposeOn more coherent and natural. In addition, P6 points out the stability of the tempo: “The music written by Suno sometimes accelerates or decelerates suddenly, whereas ComposeOn's tempo is more stable.” This is further evidence of ComposeOn's strength in maintaining musical consistency.

Finally, ComposeOn shows more variation and innovation in its music writing, as articulated in P5's review, “Suno continues music that feels like it's repeating the same chords and melodies without advancing or developing; ComposeOn at least gives me a sense that the music is evolving because it doesn't all sound too much like the previous input.ComposeOn at least gives me ComposeOn at least gives me a sense that the music is evolving, as it doesn't all sound too much like the previous input. It demonstrates ComposeOn's ability to introduce new elements while maintaining musical coherence, making the music more layered and diverse.”

\subsection{ComposeOn increased participants' willingness and confidence to create music}

Our findings indicate that ComposeOn significantly enhanced participants' willingness and confidence in music creation. This improvement can be attributed to two key features of the system: high visualization and interactivity, and high explainability.

\subsubsection{High Visualization and Interactivity}

ComposeOn's interface provides a highly visual and interactive experience, which proved particularly beneficial for novice users. The system displays both user input and AI-recommended melodies on a traditional musical staff, offering real-time visual cues during playback. This feature allows users, especially beginners, to gain a deeper understanding of musical structure, including how notes form chords and how chord progressions work.

The system's interactivity is evident in its user-guided recommendation process. Participants can control the flow of suggestions using the "continue" and "end" buttons, giving them agency over the composition process. Furthermore, ComposeOn offers contextually appropriate chord and rhythm options for each measure, allowing users to experiment with different musical elements. For instance, one participant reported successfully changing a 1-4-5-1 chord progression to a 1-2-5-1 progression, demonstrating the system's flexibility and its capacity to facilitate learning through experimentation.

\subsubsection{High Explainability}

ComposeOn's high explainability feature significantly contributes to users' understanding and confidence. The system provides detailed explanations for its musical continuations, covering aspects such as chord progression, rhythm, and ornamental notes. These explanations are tailored to three proficiency levels - beginner, intermediate, and expert - ensuring that users receive information appropriate to their knowledge level.

The granularity of explanations is noteworthy; users can request explanations for individual notes, measures, or entire phrases. This feature not only clarifies the system's recommendations but also reinforces fundamental musical concepts. As one participant noted, "This feature not only helped explain why certain recommendations were made, but it also helped me remember basic information like chord composition more clearly. I could associate the sound of a chord in a phrase with its composition, an experience that's hard to achieve when learning chords in isolation."

Additionally, the integrated "music theory mentor" feature provides quick access to theoretical knowledge, offering suggested terms and concepts like the circle of fifths. This feature's effectiveness was demonstrated when a participant (P1) first encountered an explanation of the I-VI-V-I chord progression and then used the music theory mentor to gain a deeper understanding of why this progression is harmonically pleasing.

In conclusion, ComposeOn's combination of high visualization, interactivity, and explainability creates a supportive environment for music creation. By demystifying the composition process and providing immediate, context-specific feedback, the system empowers users to engage more confidently with music creation, regardless of their initial skill level. This approach not only facilitates the creation of music but also enhances users' overall musical understanding, potentially contributing to long-term skill development and musical appreciation.

\section{Discussion}\label{sec:Discussion}
\subsection{Generative and Music Theory-Based Composition}

The emergence of generative AI music tools like Suno has transformed digital music creation. These tools use deep learning algorithms to generate complex, multi-layered compositions with minimal user input ~\cite{briot2020deep}. They excel at creating diverse musical textures, particularly for background music or ambient sounds. However, our findings align with recent research indicating that while generative tools may produce impressive initial results, they often lack the musical coherence and logical structure many listeners expect, especially upon repeated listening ~\cite{carnovalini2020computational,r22}.

The challenges faced by generative AI in music creation stem from fundamental differences between music and language structures. While AI has made significant progress in natural language processing and generation, music composition presents unique challenges. As Herremans et al. (2017) point out, musical patterns, particularly in chord progressions and melodic structures, are more constrained and rule-bound than language patterns ~\cite{herremans2017functional,r13}. This inherent structure in music makes it crucial for composers, especially beginners, to have a basic understanding of these musical rules and structures.

Compared to generative approaches, music theory-based tools like ComposeOn are better suited for users who want to understand and actively participate in the composition process. These tools are particularly beneficial for beginners and intermediate learners who wish to develop musical skills while creating. ComposeOn's structured approach follows established music theory principles, which may produce simpler but more coherent and logically clear compositions. This approach aligns with pedagogical research emphasizing the importance of active learning in music education ~\cite{wright2010informal,paule2017music}.

Recent studies highlight the potential of theory-based composition tools in enhancing musical understanding and creativity. For example, Paule-Ruiz et al. (2017) found that students using theory-based composition software showed significant improvements in understanding musical concepts and creating original compositions ~\cite{paule2017music}. This supports the view that tools like ComposeOn can serve not only as composition aids but also as effective learning platforms.

While generative AI tools have opened new possibilities for music creation, our analysis suggests that theory-based approaches retain important value, particularly for educational purposes and those seeking to create more structured compositions. As the field evolves, we may see increasing convergence of these approaches, potentially leading to tools that offer both the creative freedom of generative AI and the guiding structure of music theory.

\subsection{The Importance of Melody Input for Beginners}

ComposeOn's melody input feature plays a crucial role in making music creation accessible to beginners. By allowing users to input their melodic ideas, whether through singing or simple note input, ComposeOn bridges the gap between musical inspiration and realization. This feature addresses a key finding from our preliminary research: many novice composers have melodic ideas but lack the technical knowledge to develop them ~\cite{r14,r15}.

Melody's input serves as a starting point, providing the system with a framework to build upon. It allows users to see their ideas transformed into structured musical works, enhancing their sense of ownership and creative engagement. Moreover, this feature helps users understand how their initial ideas fit into larger musical structures, promoting learning about harmony, rhythm, and form in a practical, hands-on way ~\cite{r19}.

By using user-provided melodies as the basis for composition, ComposeOn also ensures that the final product retains personal character, addressing concerns about AI-generated music lacking individual creativity ~\cite{r20}. This approach strikes a balance between AI assistance and user creativity, particularly suitable for beginners who want to learn while creating.

The importance of melody input for novice composers has been recognized in various music creation systems. For instance, the MusicMaker system developed by Chuan and Chew ~\cite{chuan2007} allows users to input melodies via MIDI keyboard or singing, which are then automatically harmonized. Similarly, the Hyperscore system ~\cite{farbood2004} enables users to draw melodic contours that are then converted into musical phrases, making composition accessible to those without formal musical training ~\cite{r18}.

In the mobile app domain, SongSmith ~\cite{simon2008} and Bean Academy ~\cite{r16} pioneered the approach of inputting melodies through singing, automatically generating accompaniments based on the user's vocal input. More recently, AI-driven systems like AIVA ~\cite{hadjeres2017} have also incorporated melody input features, allowing users to provide initial ideas that are then expanded by AI.

These examples from the literature highlight the widespread recognition of melody input as a key feature in making music creation more accessible to beginners. ComposeOn builds on this established principle, combining it with advanced AI capabilities to provide a comprehensive learning and creation environment for novice composers.

\subsection{The Role of Explainability in Music Learning}

A significant advantage of ComposeOn over other music creation and learning tools is its high level of explainability. This feature not only functionally distinguishes ComposeOn from other generative AI tools but, more importantly, transforms the system into an interactive music learning environment. Explainability has become increasingly important in AI and machine learning, especially in educational and creative applications ~\cite{gunning2019,r17}. ComposeOn applies this trend to music education, pioneering a new paradigm of learning.

ComposeOn's explainability features serve a dual purpose: they help users understand the composition process while teaching them music theory. This approach closely aligns with modern teaching theories, particularly those emphasizing learning by doing and understanding the reasons behind rules and conventions ~\cite{kolb2014experiential}. By providing detailed, context-specific explanations, ComposeOn enables users to see the decisions made during the composition process and understand the musical logic behind these decisions. This transparency not only enhances users' understanding of music creation but also cultivates their critical thinking skills, enabling a deeper comprehension of musical structures and theory.

Another key advantage of ComposeOn is its multi-tiered explanation feature (beginner, intermediate, expert). This design allows the system to remain relevant as users' musical knowledge progresses, enabling ComposeOn to serve as a long-term companion for music learning and composition. This approach aligns with cognitive load theory, which emphasizes that learning materials should be adjusted according to the learner's expertise level to optimize learning outcomes ~\cite{sweller2019cognitive}. By providing explanations tailored to the user's current level, ComposeOn ensures that the learning process is challenging but not overwhelming, creating an ideal learning environment.

Furthermore, ComposeOn's explainability features not only impart technical knowledge but also foster users' musical intuition and creativity. By explaining why certain musical choices work well, the system helps users develop their own creative judgment. This approach supports constructivist learning theory, which posits that learners actively construct knowledge by combining new information with existing knowledge structures ~\cite{ormrod2020human}. In ComposeOn, users are not passive recipients of information but active participants in the knowledge construction process, contributing to deeper and more lasting learning ~\cite{r21}.

ComposeOn also effectively bridges the gap between music theory and practice through its explainability features. By explaining theoretical concepts in the context of actual composition, users can immediately see the application of these concepts. This immediate feedback and application helps deepen understanding and significantly improves knowledge retention ~\cite{hattie2007power}. By connecting abstract music theory concepts with concrete music creation practices, ComposeOn creates a comprehensive learning environment that facilitates the translation of theoretical knowledge into practical skills.

The multi-tiered explanation system of ComposeOn allows users to customize their learning experience according to their needs and interests. This personalized approach aligns with self-determination theory, which emphasizes learners' agency and autonomy in the learning process ~\cite{ryan2020self}. By allowing users to choose the complexity of explanations, ComposeOn empowers learners to control their own learning pace, enhancing learning motivation and engagement.

ComposeOn's explainability features not only enhance its value as a composition tool but also transform it into a powerful music learning platform. By seamlessly integrating theoretical knowledge with practical application, providing personalized learning experiences, and fostering critical thinking and creativity, ComposeOn offers users a comprehensive music education environment. This innovative approach not only supports long-term learning and creation processes from beginner to expert but also has the potential to revolutionize how music education is delivered, making the learning process more interactive, personalized, and effective.

\section{Conclusion}\label{sec:Conclusion}
ComposeOn is not merely an innovative tool for music composition; it represents a shift in the intersection of music education and technology. Through its user-friendly design and multi-level theoretical guidance, ComposeOn effectively lowers the barriers traditionally associated with music creation, making the process accessible to those without formal musical training. For users lacking a background in music theory, ComposeOn offers a platform that fosters incremental learning, enabling them to transform simple melodic ideas into fully developed compositions through interactive guidance.

Our findings indicate that ComposeOn demonstrates clear superiority over generative AI-based tools such as Suno in terms of musical structure and coherence. While Suno may generate sonically complex and multi-layered compositions, its output often lacks logical progression and structural integrity, especially upon repeated listening. In contrast, ComposeOn produces music that is not only more consistent and cohesive but also aligns with established principles of music theory. Moreover, the high level of explainability embedded in ComposeOn significantly enhances users' understanding of fundamental concepts such as chord progressions and rhythmic patterns, blending theory with practice to provide a more immersive and effective learning experience.

A key strength of ComposeOn lies in its ability to elevate users' confidence and motivation to create music. Through its highly visual and interactive interface, users can actively engage with the composition process, taking control of melodic development and exploring various harmonic and rhythmic options. The system’s flexibility allows for real-time adjustments, empowering users to make informed creative decisions. This interactive approach not only increases user engagement but also transforms them from passive recipients of AI-generated music into active composers with a deeper sense of creative ownership.

Furthermore, ComposeOn’s multi-tiered explanation system caters to users of varying skill levels, offering appropriate theoretical insights at each stage of their development. This layered approach ensures that users, whether beginners or intermediate learners, can progress at their own pace, reinforcing long-term knowledge retention through the immediate application of theoretical concepts. The integration of music theory into the creative process bridges the gap between abstract learning and practical application, thereby enhancing the overall educational value of the platform.


\bibliographystyle{ACM-Reference-Format}
\bibliography{references}

\end{document}